\documentclass[aps,prl,preprint,groupedaddress]{revtex4-1}

\usepackage{graphicx}
\usepackage{amssymb}
\usepackage{amsmath}

\DeclareRobustCommand{\Chi}{{\mathpalette\irchi\relax}}
\newcommand{\irchi}[2]{\raisebox{\depth}{$#1\chi$}} 

\begin{document}

\title{Charge Order from Orbital-dependent Coupling Evidenced by NbSe$_2$}

\author{Felix Flicker}
  \email{flicker@physics.org}
  \affiliation{University of Bristol, H. H. Wills Physics Laboratory, Tyndall Avenue, Bristol, BS8 1TL, UK}
\author{Jasper van Wezel}
  \email{vanwezel@uva.nl}
  \affiliation{University of Bristol, H. H. Wills Physics Laboratory, Tyndall Avenue, Bristol, BS8 1TL, UK}
  \affiliation{Institute for Theoretical Physics, University of Amsterdam, 1090 GL Amsterdam, The Netherlands}

\begin{abstract}
Niobium diselenide has long served as a prototype of two-dimensional charge ordering, believed to arise from an instability of the 
electronic structure analogous to the one-dimensional Peierls mechanism. Despite this, various anomalous properties have recently been identified 
experimentally which cannot be explained by Peierls-like weak-coupling theories. 
We consider instead a model with strong electron-phonon coupling, taking into account both the full momentum and orbital dependence of the 
coupling matrix elements. We show that both are necessary for a consistent description of the full range of experimental observations. 
We argue that NbSe$_2$ is typical in this sense, and that any charge-ordered material in more than one dimension will generically be shaped by the 
momentum and orbital dependence of its electron-phonon coupling as well as its electronic structure. The consequences will be observable in 
many charge-ordered materials, including cuprate superconductors.

\end{abstract}

\maketitle

The Charge Density Wave (CDW) order in $2H$-NbSe$_2$ has been surrounded by controversy since its first observation nearly 40 years ago~\cite{WilsonEA74}. 
The most basic unresolved question concerns the driving mechanism of the ordering transition. In quasi-1D materials, it is well known 
that charge order arises from a Peierls instability: any interaction between states at $E_F$ results in a periodic modulation of both the electronic charge 
density and the atomic lattice positions, which lowers the free energy through the corresponding opening of a band gap~\cite{Peierls}. Such a mechanism 
is termed `weak coupling' since an arbitrarily small interaction will drive the system to order at low enough temperatures.
In higher-dimensional systems, a nesting of the Fermi surface is typically required for such an instability to be effective. In NbSe$_2$, however, Fermi 
surface nesting does not occur~\cite{WilsonEA74,Doran78,BorisenkoEA09,JohannesEA06}. The lack of a lock-in transition and the coexistence of CDW order 
with superconductivity below $7.2\,$K nonetheless seem to suggest a weak-coupling origin of the charge order. As a result a number of alternative weak-coupling 
mechanisms have been proposed, based on nested saddle-points in the electronic 
dispersion~\cite{RiceScott75}, local field effects~\cite{JohannesEA06}, or a combination of weak nesting with momentum-dependent electron-phonon 
coupling~\cite{Doran78,WeberEA11}.

There are a number of experimental observations, however, which are hard to reconcile with weak-coupling approaches in NbSe$_2$. The first concerns the size of 
the electronic gap in the charge-ordered state. Kinks in the Density Of States (DOS) observed by planar 
tunnelling experiments have been interpreted to arise from a gap of $\Delta = 35\,$meV~\cite{HessEA91,WangEA90}, while only much smaller gaps between $2-5\,$meV 
were seen by high-precision Angle Resolved Photoemission Spectroscopy (ARPES) experiments~\cite{BorisenkoEA09,RahnEA12}, and older studies even reported 
no gap at all in ARPES and resistivity measurements~\cite{StraubEA99,HarperEA75}. Assuming a weakly-coupled driving mechanism for the charge order, 
the size of the gap is expected to be proportional to the transition temperature, $\Delta \simeq 1.76 k_B T_C $, in direct analogy to the BCS theory of 
superconductivity~\cite{VarmaSimons83}. A gap size of $35\,$meV would then be far too large to explain the observed $T_{\textrm{CDW}}$ of 
$33\,$K~\cite{MonctonEA75}. The $5\,$meV gap consistent with ARPES data is of the expected size, but it occurs only in isolated points on the Fermi 
surface~\cite{BorisenkoEA09,RahnEA12}, raising the question of how the correspondingly small gain in electronic energy can overcome the cost of introducing the 
periodic lattice distortions associated with CDW order. Additionally, while the gaps seen in ARPES measurements are connected by the CDW wave vector, other 
equally well-connected points support either only a much smaller gap, or no gap at all~\cite{BorisenkoEA09,RahnEA12}. 
It was recently suggested, based on Scanning Tunneling Spectroscopy (STS) measurements~\cite{SoumyanarayananEA13}, that a particle-hole asymmetric gap of 
$12\,$meV, centred above $E_F$, exists in NbSe$_2$. The offset in energy explains why ARPES only sees a smaller gap size, while the value of $12\,$meV is 
consistent with the order of magnitude expected for a strongly-coupled CDW, which may have $\Delta \sim 4 k_B T_C $ in analogy to strong-coupling 
superconductors~\cite{RainerEA92,WangEA07,HashimotoEA14}.

The idea that the CDW order in NbSe$_2$ is of the strong coupling variety, in which a sizable electron-phonon coupling is essential in causing an instability 
of the electronic structure, is also consistent with two further recent experimental observations. Firstly, it was observed in soft X-ray scattering 
experiments that at the transition temperature a broad momentum-range of phonons simultaneously softens towards zero energy~\cite{WeberEA11,WeberEA13}. This is 
in stark contrast to the sharp Kohn anomaly expected in a Peierls-like transition~\cite{Gruner88}, and may be understood to be an effect of the importance of 
the entropy carried by phonon fluctuations in the strongly-coupled scenario~\cite{McMillan75,YoshiyamaEA86}. These phonon fluctuations are localized in real 
space, and therefore necessarily involve a broad range of momenta. In an electronically nested system, on the other hand, the nesting vector localizes 
any fluctuations to a single point in momentum space. The second experimental result arguing for a strongly-coupled transition is the observation of a 
reduction of the density of states persisting well above the transition temperature, in what has been termed a pseudogap 
regime~\cite{BorisenkoEA09,ChatterjeeEA14,ArguelloEA14}. In strongly coupled CDW materials, a pseudogap typically appears through the presence of locally 
fluctuating charge order without long-range coherence~\cite{ChatterjeeEA14}. These same charge fluctuations have recently been suggested to play a central 
r\^{o}le in the description of the pseudogap phase of cuprate superconductors~\cite{EmeryEA99,WalstedtEA11,ChangEA12,GhiringhelliEA12}.
The precise nature of the pseudogap in these materials, however, is difficult to establish owing to the presence of various nearby competing orders. 
NbSe$_2$, being free of such complications, provides a test bed on which the formation and characteristics of a pseudogap phase due to charge 
fluctuations can be studied directly.

Here we present a theoretical analysis of the charge order in NbSe$_2$ based on a model of strong electron-phonon coupling. 
Although the transition is phonon driven we find that a full knowledge of the electron states scattered from and to, including both wave vector and orbital 
dependence, is crucial in explaining the experimental data. Despite a lack of nesting, the limited matching of states at $E_F$ acts in unison with the momentum 
dependence of the electron-phonon coupling to select out the CDW ordering vector, while the orbital characters of the bands naturally explain why a CDW gap 
appears primarily in only one of the Fermi surface pockets. 
The resulting momentum dependence of the gap itself agrees with ARPES experiments as well as with the observed particle-hole asymmetry in the DOS.
Finally we find that the phonon modes are softened over a broad range of wave vectors, and that including the presence of strong phonon fluctuations leads to a 
suppression of the transition temperature, implying a range of intermediate temperatures dominated by incoherent, fluctuating, charge order, and a 
corresponding pseudogap.
In light of these findings, and the generic nature of the electronic structure of NbSe$_2$, we argue that the momentum and orbital dependence of the 
electron-phonon coupling may be expected to play a similarly central part in the description of other charge ordered materials, especially in the presence of 
relatively strong coupling as in the case of the layered high-$T_c$ superconductors.

\section*{Results}

\subsection*{The extent of electronic nesting}

\begin{figure}  
  \includegraphics[width=0.7\columnwidth]{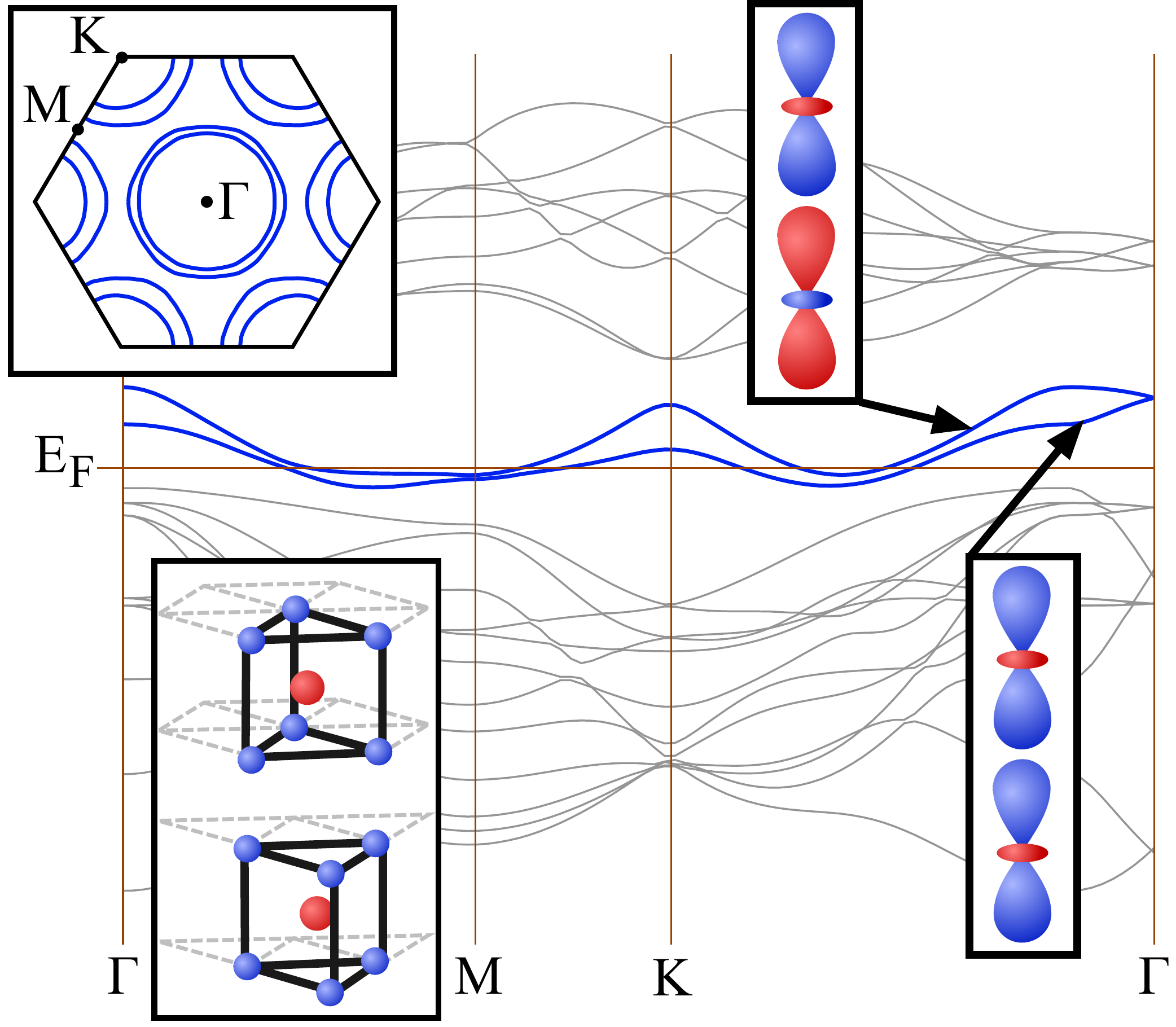}  
  \caption{\label{bandstructure}
  The band structure of NbSe$_2$ modelled by a tight-binding fit to both ARPES data and LDA calculations. The two bands crossing $E_F$ (thick lines) are the ones involved in the formation of charge order. They consist of bonding and anti-bonding combinations of the two Niobium $d_{3z^2-r^2}$ orbitals within a unit cell, as indicated. \emph{Upper inset:} the Fermi surface resulting from the two bands crossing $E_F$. \emph{Lower inset:} the layered atomic structure of NbSe$_2$, with Nb in blue and Se in red. The unit cell, containing two formula units across consecutive sandwich layers, is indicated.
  }
  \end{figure}

Niobium diselenide is a layered material, in which hexagonal layers of Niobium atoms are sandwiched between similar layers of Selenium atoms, displaced so that 
they lie above and below half of the Nb interstitial locations, as shown in the inset of Fig.~\ref{bandstructure}. Consecutive sandwich layers are displaced 
to have the complementary half of the 
interstices occupied, giving two formula units per unit cell. The Fermi surface consists of two concentric barrel-shaped pockets centred around both the 
$\Gamma$ and the $K$ points, as well as a very small pancake-shaped pocket surrounding $\Gamma$, which we ignore in the analysis below.
The inner pockets at $K$ are seen in ARPES to develop the largest CDW gaps~\cite{BorisenkoEA09,RahnEA12}. The CDW itself is of the $3Q$ type, with three 
equivalent superposed modulations at $120^{\circ}$ with respect to one other.

To describe the electronic states, we employ a Slater-Koster 
tight-binding fit to the band structure using a basis of Se-$4p$ and Nb-$4d$ orbitals~\cite{KosterSlater54,DoranEA78}. 
For the two main bands crossing $E_F$ that give rise to the barrel-shaped pockets (which we will refer to as inner, developing the CDW instability, and 
outer), the fit is constrained by data from ARPES measurements~\cite{RahnEA12}, while the remaining $20$ bands are fit to recent LDA 
calculations~\cite{JohannesEA06}. 
In agreement with earlier reports~\cite{DoranEA78}, we find that both the inner and outer bands at $E_F$ are  primarily composed of the $d_{3z^2-r^2}$ orbitals 
centred on the two Nb atoms within a unit cell, in the form of bonding (inner) and anti-bonding (outer) combinations (as shown in Fig.~\ref{bandstructure}). 

Focussing on the two non-interacting bands at $E_F$, we introduce the interaction between electrons $\psi$ and phonons $\varphi$:
\begin{align}\label{H_int}
\hat{H}_{int}=\sum_{kq}g_{\mathbf{k},\mathbf{k}+\mathbf{q}}^{m, n} ~ \hat{\varphi}^{\phantom{*}}_{q} \hat{\psi}_{\phantom{\dagger} k}^{\dagger m} 
\hat{\psi}_{k+q}^{n},
\end{align}
where the bare phonon dispersion can be described by a Brillouin function tending to a maximum of $11.2\,\textrm{meV}$ at the zone 
boundary~\cite{WeberEA13}. 
The electron-phonon coupling $g$ depends on the momenta, as well as the band indices $m$ and $n$, of the the ingoing and outgoing 
electron states. Following Varma \emph{et al.} we model both of these dependencies using knowledge purely of the electronic structure, which has been shown 
to work well for a range of transition metals and their compounds~\cite{VarmaEA79}. In terms of the matrices of tight-binding overlap integrals 
$S_{\mathbf{k}}$ and generalized eigenvectors $A_{\mathbf{k}}$, the electron-phonon coupling matrix elements are given by~\cite{VarmaEA79}:
\begin{align}\label{g_full}
\mathbf{g}_{\mathbf{k}\mathbf{k}'}^{mn}=\mathbf{v}_{\mathbf{k}}^{m}\left[A_{\mathbf{k}}^{\dagger}S_{\mathbf{k}}A_{\mathbf{k}'}\right]^{mn}-\left[A_{
\mathbf{k}}^{\dagger}S_{\mathbf{k}'}A_{\mathbf{k}'}\right]^{mn}\mathbf{v}_{\mathbf{k}'}^{n},
\end{align}
where a constant prefactor has been omitted, and the band velocity $\mathbf{v}^n$ for electrons in band $n$ is defined in terms of the band energy 
$\xi_{\mathbf{k}}^n$ as $\mathbf{v}^n_\mathbf{k}=\partial\xi_{\mathbf{k}}^{n}/\partial\mathbf{k}$. In the interaction of equation~\eqref{H_int} we restrict 
attention to the longitudinal component of $\mathbf{g}$, since only the longitudinal acoustic phonons are observed to soften in inelastic neutron and X-ray 
scattering experiments~\cite{WeberEA13}.

Inserting the tight binding results into equation~\eqref{g_full}, we find that the inter-band electron-phonon coupling is strictly zero, 
while the intra-band coupling is about a factor of three stronger for the inner band than for the outer. This result immediately 
explains why the experimentally observed CDW gap is so much more pronounced in the inner band~\cite{BorisenkoEA09,RahnEA12}, since the static electronic 
susceptibility scales with the square of the electron-phonon coupling:
\begin{align}\label{D2}
\Chi_{0}\left(\mathbf{q}\right) \propto \sum_{\mathbf{k}}|g^{mn}_{\mathbf{k},\mathbf{k}+\mathbf{q}}|^2\frac{f(\xi^m_{\mathbf{k} 
\vphantom{\mathbf{q}}})-f(\xi^n_{\mathbf{k}+\mathbf {q}})}{\xi^n_{\mathbf{k}+\mathbf{q}} - \xi^m_{\mathbf{k}}}.
\end{align}
Here $f\left(E\right)$ is the Fermi-Dirac distribution function, and the static Lindhard function is given by $\Chi_{0}$ when 
$g=1$. The absence of interband coupling causes the contributions of the two bands to the susceptibility to be entirely independent. This means that the 
lower value of the intra-band coupling in the outer band leads to an order of magnitude suppression of its contribution to $\Chi_{0}$, and therefore 
to a correspondingly reduced gap in its Fermi surface pocket. It is thus the orbital character of the electronic bands, and consequently the orbital dependence 
of the electron-phonon coupling, that causes a relative size difference between the CDW gaps in the pockets around the $K$-point, even if they are equally well 
nested.

\begin{figure}  
  \includegraphics[width=0.75\columnwidth]{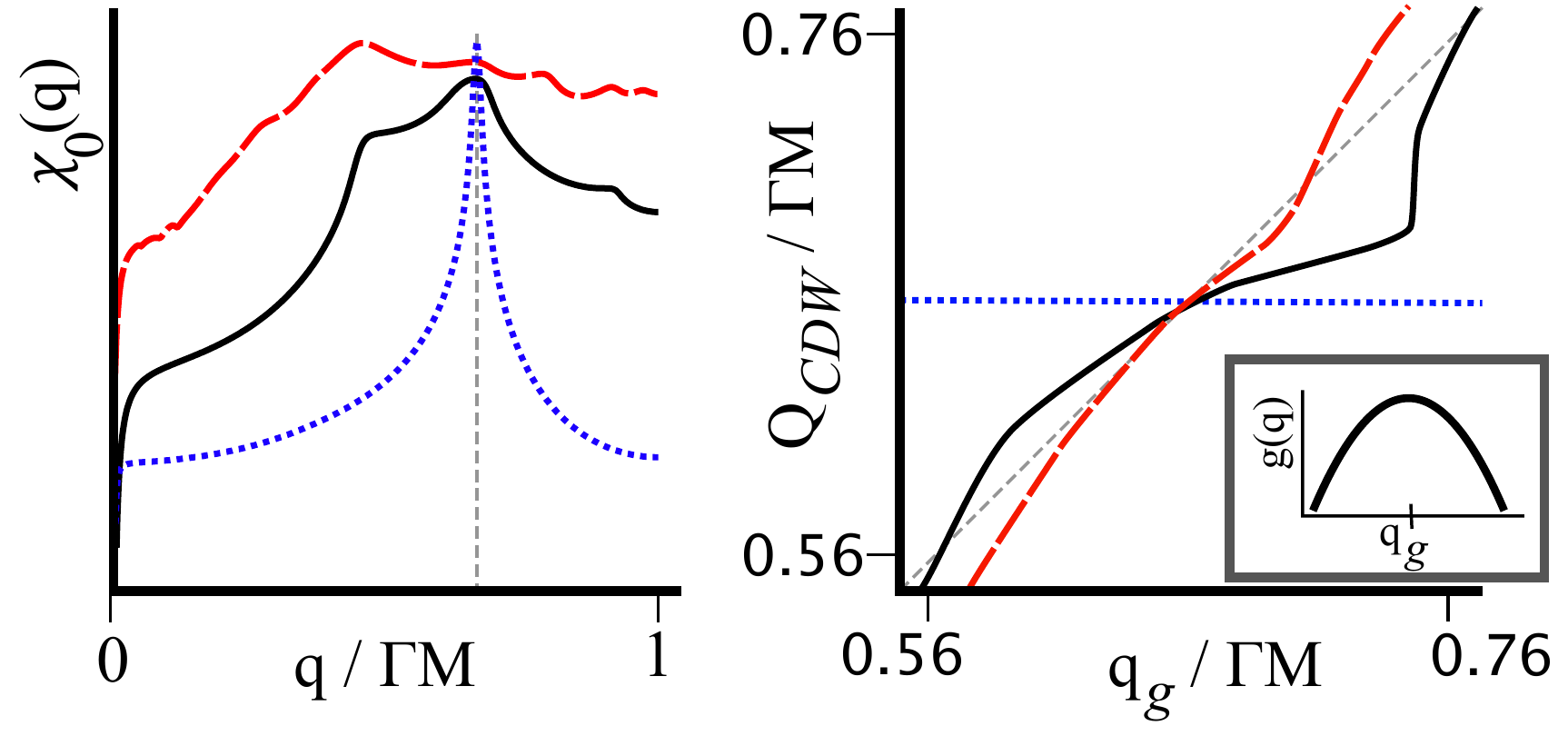}  
  \caption{\label{nesting}
\emph{Left:}  electronic susceptibility as a function of momentum transfer. Blue dotted line: in a perfectly nested 1D band structure, the susceptibility is sharply peaked at $\mathbf{q} = \mathbf{Q}_{CDW}$. 
Red dashed curve: the result for the two bands in NbSe$_2$ crossing $E_F$, and an orbital-independent electron-phonon coupling. The susceptibility is nearly flat, without any peaks. Black solid curve: the susceptibility of just the inner band, in which a CDW gap is observed. Although a clear maximum can be seen at $\mathbf{Q}_{CDW}$, it is not a sharp divergence indicative of perfect nesting.
\emph{Right:} evolution of the CDW vector $\mathbf{Q}_{CDW}$ with varying peak position $\mathbf{q}_{g}$ of the model electron-phonon coupling (see inset). Blue 
dotted line: in a perfectly nested 1D band structure the electronic structure determines the CDW formation and $\mathbf{Q}_{CDW}$ is independent of $\mathbf{q}_{g}$. 
Red dashed curve: the result for an orbital-independent electron-phonon coupling and the two bands crossing $E_F$, which is close to a straight line at 
$45^{\circ}$, indicative of a negligible 
r\^{o}le of the electronic structure. Black solid curve: the evolution relevant for NbSe$_2$, given by just the inner band as imposed by the orbital-dependent 
electron-phonon coupling. This curve interpolates between the flat plateau of a perfectly nested structure, and the $45^{\circ}$ line dominated by just the structure of the electron-phonon coupling.}
  \end{figure}

The expression for the susceptibility in equation~\eqref{D2} can further be used to quantify both the amount of nesting present in NbSe$_2$, and to evaluate 
the relative contribution of the electron-phonon coupling matrix elements in selecting the observed value of the (incommensurate) ordering wave 
vector $\mathbf{Q}_{CDW}=\left(1-\delta\right)\frac{2}{3}\,{\Gamma M}$, with $\delta\approx0.02$~\cite{WilsonEA74}. In the left panel of 
Fig.~\ref{nesting}, three susceptibilities are compared. The dotted blue line shows the susceptibility for a perfectly nested model with a single 1D band of 
cosine dispersion. The sharp peak characteristic of a nested Fermi surface is apparent. The dashed red line is the bare susceptibility in NbSe$_2$, 
obtained by setting $g=1$ in equation~\eqref{D2}. This does not include any of the momentum and orbital dependence of the electron-phonon coupling, and is 
the form that has been widely used in previous studies~\cite{SoumyanarayananEA13,JohannesEA06,RahnEA12}. It is almost completely flat except 
for a very broad and low maximum far from $\mathbf{Q}_{CDW}$, and can therefore not explain the observed ordering vector. The solid black line, on the other 
hand, displays the susceptibility of just the inner band. A broad but distinct maximum develops at the experimentally observed CDW vector. The selection of the 
ordering vector is thus shown to arise from the combined influence of both the electronic structure and the orbital and momentum dependent electron-phonon 
coupling, with neither contribution being negligible~\cite{BorisenkoEA09,RahnEA12}.

Restricting attention to the inner band, we can quantify the extent to which nesting contributes to the selection of $\mathbf{Q}_{CDW}$. For this 
purpose we temporarily switch to a simplified form of $g$, which is a 
function only of the momentum transfer: $g = -a \left(q_{g}-q\right)^2+g_{max}$. Even though in general the shape of the susceptibility is 
determined by both the incoming momentum and the momentum transfer, modelling the electron-phonon coupling in this simplified form allows us to 
directly quantify the effect of varying the position of its peak, and thus its influence on the position of the peak in $\Chi_{0}$.
The values of the parameters $a$, $q_{g}$, and $g_{max}$ can be estimated by fitting the experimentally observed phonon softening~\cite{WeberEA11} to the 
expression for the renormalized phonon energy in the Random Phase Approximation (RPA), $\Omega_{\textrm{RPA}}^2 = 
\Omega_0 \left(\Omega_0 - \Chi_{0}\right)$, where we take $\Omega_0$ to be the high temperature (unrenormalized) dispersion. In the right panel of Fig.~\ref{nesting} 
we plot the variation of the maximum in the susceptibility (\emph{i.e.} the expected $\mathbf{Q}_{CDW}$) as $\mathbf{q}_{g}$ is varied along $\Gamma M$). Again, the 
blue dotted line represents the situation of a perfectly nested band structure, in which the electron-phonon coupling never overcomes the peak in the 
susceptibility, and thus has no effect on $\mathbf{Q}_{CDW}$. The red dashed line has the same form of $g$ for all bands, and is close to a straight line 
at $45^{\circ}$ (grey dashed line). A true $45^{\circ}$ line would indicate that $\mathbf{Q}_{CDW}=\mathbf{q}_{g}$, implying that the CDW vector is 
completely determined by the electron-phonon coupling, with the electronic structure having no influence at all. The solid black line, which shows the 
behaviour of just the inner band, interpolates between these two extremes. There is no plateau, but there is an indication of some flattening, suggesting that 
while it is mainly the momentum dependence of the electron-phonon coupling which determines the value of CDW ordering vector, the r\^{o}le of the electronic 
structure cannot be neglected entirely.

\subsection*{The CDW Gap}

\begin{figure}
  \includegraphics[width=0.9\columnwidth]{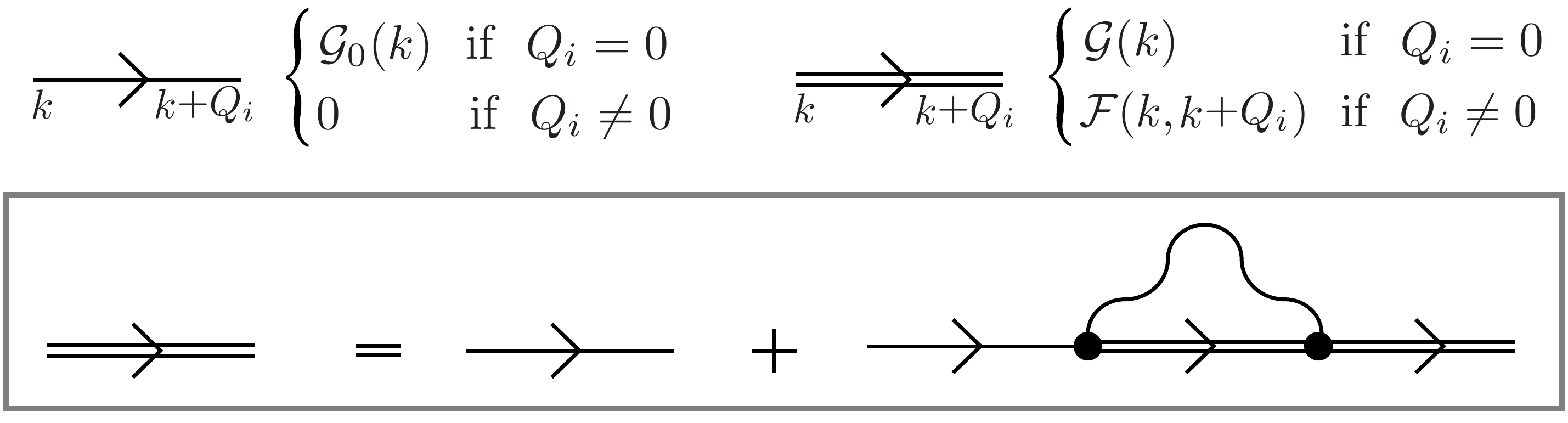}  
  \caption{\label{diagrams}
Diagrammatic form of the self-consistent RPA gap equations. The single arrows indicate bare electronic Green's functions $\mathcal{G}_0$, and the double arrows 
the renormalized Green's functions $\mathcal{G}$ and $\mathcal{F}$. The anomalous functions $\mathcal{F}(k,k+Q_i) =  \langle \hat{\psi}_{k} \hat{\psi}_{k+Q_i}^{\dagger} \rangle$ do not conserve crystal momentum, and correspond to the CDW order parameter. Notice that the final diagram includes a summation over the anomalous momentum $Q_i$ of the internal electronic propagator.
  }
  \end{figure}

Having identified the importance of both electron-phonon coupling and electronic structure, we now revert to using the full expression of 
equation~\eqref{g_full}. To compare the size and shape of the CDW gap resulting from the 
orbital and momentum dependent $g$ to the experimentally observed behaviour, the RPA gap equations shown diagrammatically in Fig.~\ref{diagrams} are solved 
self-consistently.
We simplify the calculation by using the $9\times9$ gap matrix which includes all higher harmonics appropriate for a commensurate CDW wave vector at $\mathbf{Q}=\frac{2}{3}\,{\Gamma M}$, as an approximation for the gap matrix of the incommensurate CDW in NbSe$_2$ with $\mathbf{Q}_{\textrm{CDW}} \simeq 0.98 \frac{2}{3}\,{\Gamma M}$.
We further fix the values of the electronic self-energies, and constrain the momentum dependence of the gap function to a five parameter tight-binding fit 
consistent with the symmetries of the lattice. A self-consistent solution for the gap function to all orders in RPA can then be found by searching for a fixed 
point in the flow of consecutive iterations of the gap equation.

\begin{figure}
  \includegraphics[width=0.9\columnwidth]{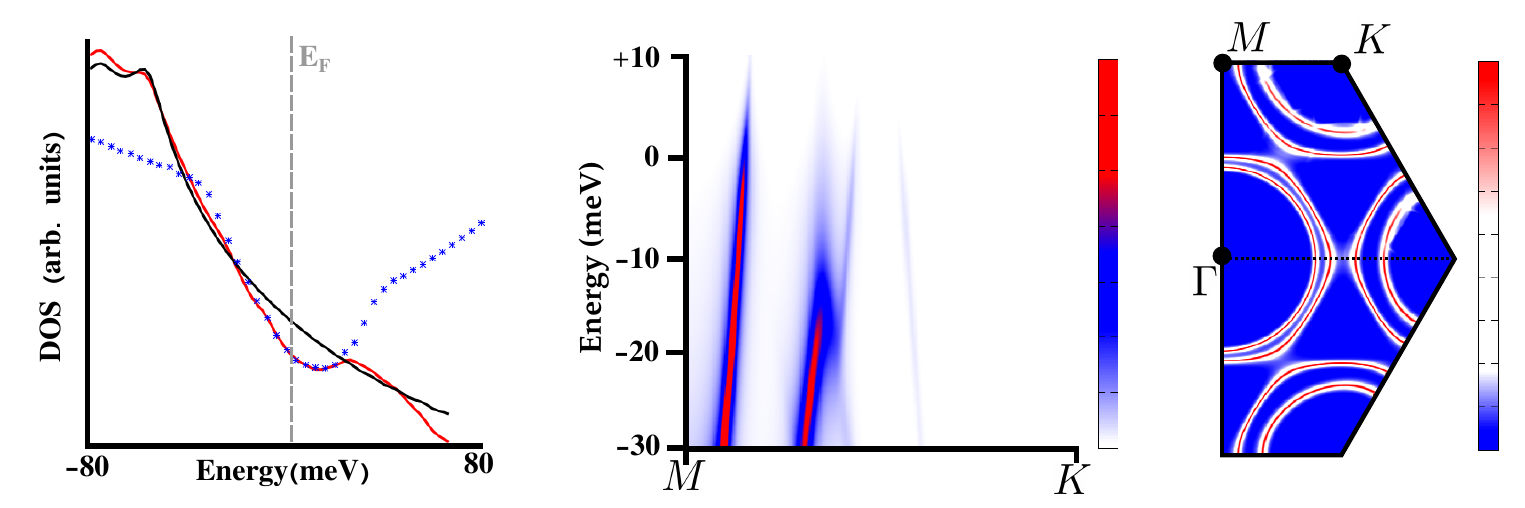}  
  \caption{\label{gap}
  \emph{Left:} the calculated density of states for the bare dispersion (black line) and renormalized dispersion (red line) compared to experimental results 
from STS measurements (blue points)~\cite{SoumyanarayananEA13}. The opening of the CDW gap leads to an asymmetric suppression of density of states, centred 
above $E_F$. \emph{Middle:} the dispersion of the calculated spectral functions along $M K$. The back-bending induced by the gap at $E_F$ is apparent in the 
band corresponding to the inner pocket about $K$. \emph{Right:} Fermi energy cut of the spectral function resulting from the self-consistent gap equations, 
showing the formation of Fermi arcs as a result of gaps opening along $M K$. Below the dashed line the bare bands are shown for comparison. Note that the outer 
band develops no discernible gap; this can be compared directly to the results of recent ARPES measurements~\cite{BorisenkoEA09,RahnEA12}.}
\end{figure}

Fig.~\ref{gap} shows energy and momentum cuts through the electronic spectral functions resulting from the solution to the gap equations, as well as their 
corresponding DOS. 
A good match with the shape of the experimentally determined DOS in STS experiments can be achieved by introducing a $4\,$meV shift in the 
chemical potential, which falls well within the range of experimental uncertainty~\cite{SoumyanarayananEA13}.
The overall strength of the electron-phonon coupling, the only free parameter in the theory, is set to give a maximum CDW gap magnitude of $\approx12\,$meV. The Fermi surface on the right of Fig.~\ref{gap} then shows a CDW gap opening on the inner band only, and Fermi arcs forming at the same locations as observed in ARPES 
measurements~\cite{BorisenkoEA09,RahnEA12,ChatterjeeEA14}.
The extent of the gap on the Fermi surface, its size (apparent from the dispersion's back-folding in 
the centre of Fig.~\ref{gap}), and its restriction to one band, are all in agreement with 
experimental findings~\cite{BorisenkoEA09,RahnEA12,ChatterjeeEA14}. The plot of the DOS on the left indicates that these results, which were previously 
interpreted to imply a gap size of $\sim 5\,$meV~\cite{BorisenkoEA09,RahnEA12}, are in fact consistent with the STS results indicating a particle-hole 
asymmetric gap of $\sim 12\,$meV~\cite{SoumyanarayananEA13,ChatterjeeEA14}. The asymmetry hides most of the gap structure above $E_F$, where photoemission 
intensity is suppressed by the Fermi-Dirac distribution~\cite{ChatterjeeEA14}. The comparison between the experimental STS data and the calculated RPA results 
shows that the particle-hole asymmetry arises naturally from the interplay between the electronic structure and the momentum-dependent electron-phonon 
coupling.

\subsection*{The Pseudogap}

\begin{figure}
  \includegraphics[width=0.5\columnwidth]{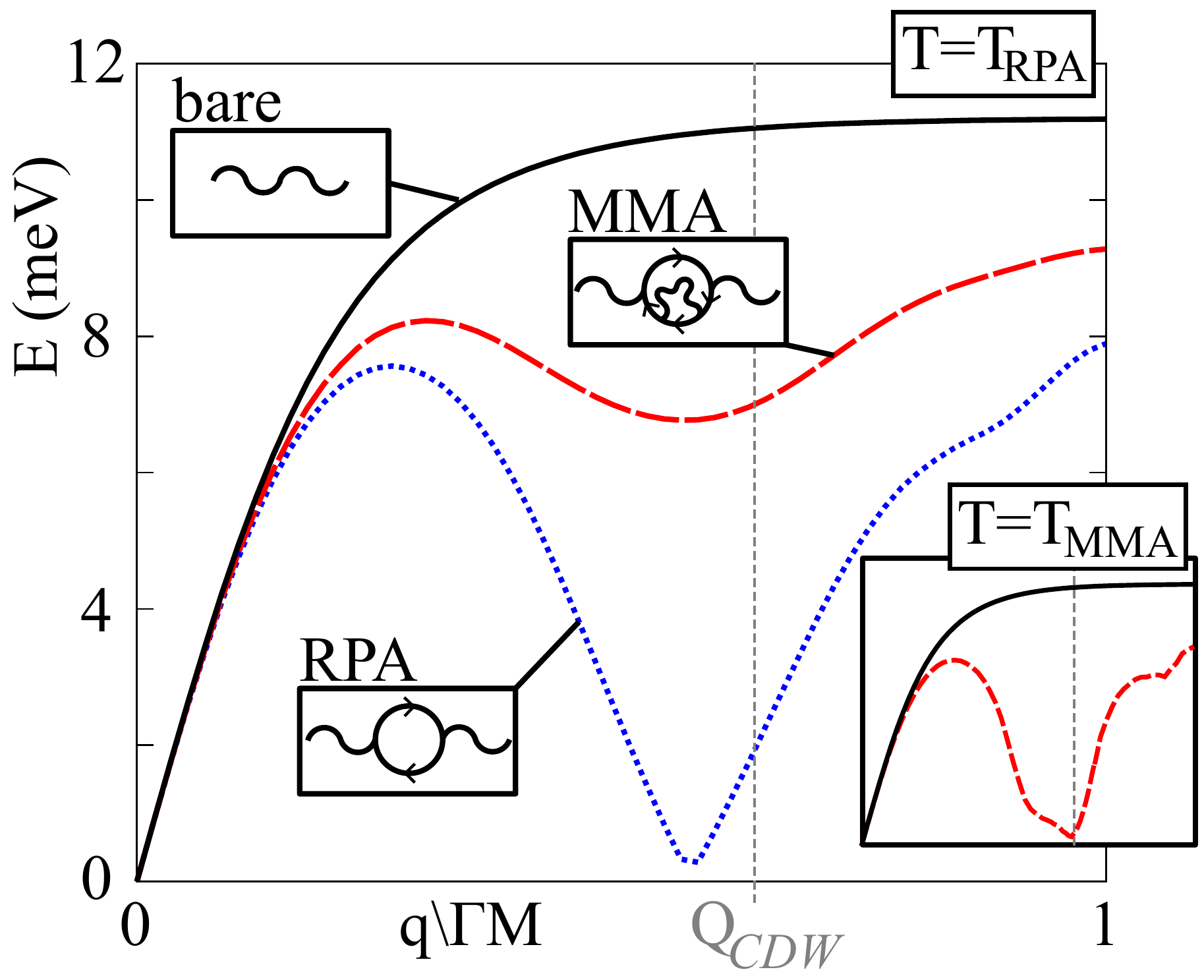}  
  \caption{\label{MMA}
  Phonon dispersion in NbSe$_2$. The solid black line indicates the bare phonon dispersion along $\Gamma M$, based on high temperature X-ray 
scattering results~\cite{WeberEA13}. The blue dotted curve shows the dispersion resulting from the (mean field) RPA calculation at 
$T_{\textrm{RPA}}=303.5\,\textrm{K}$, while the red dashed curve represents the MMA prediction (including the phonon fluctuations shown in the diagram) at the 
same temperature. The fluctuations of the phonon field can be seen to suppress the onset of charge order predicted by the RPA theory. The bottom inset shows 
the MMA dispersion at $T_{\textrm{MMA}}=33.5\,\textrm{K}$, which represents the true transition temperature. It can be seen that the momentum of the softest phonon mode 
corresponds to the experimentally observed CDW propagation vector.}
  \end{figure}

The central r\^{o}le of the orbital and momentum dependent electron-phonon coupling in determining both the CDW propagation vector and the structure of the gap, 
suggests that the charge order in NbSe$_2$ falls in the strong-coupling regime, in which the electronic structure alone is not sufficient to characterize the 
emergent charge order. In the presence of strong electron-phonon coupling, it is well-known that the entropy associated with localized lattice vibrations plays 
an essential part in the system's thermal evolution~\cite{McMillan75}. To understand the properties of NbSe$_2$ above
its CDW transition temperature, it is therefore necessary to consider the influence of phonon fluctuations beyond RPA. The so-called Mode-Mode coupling 
Approximation (MMA) includes the lowest order terms of this type, and has been shown to successfully describe the influence of 
phonon fluctuations in some of the electronic properties of related dichalcogenide materials like TaSe$_2$ and TiSe$_2$~\cite{Inglesfield80,YoshiyamaEA86}.
The additional terms included in the MMA fall into two categories: a vertex correction, which we neglect by 
appeal to Migdal's theorem because of the large momentum transfer associated with $\mathbf{Q}_{CDW}$, and self-energy corrections to the internal electron 
lines (indicated diagrammatically in Fig.~\ref{MMA}), which we include.
In agreement with recent X-ray scattering observations on NbSe$_2$~\cite{WeberEA11,WeberEA13}, the MMA-renormalized phonon spectrum shown in the inset of 
Fig.~\ref{MMA} shows a broad range of phonon modes softening near the transition temperature, rather than the sharp Kohn anomaly indicative of well-nested 
materials. The momentum of the softest phonon mode at the transition temperature coincides with the experimentally observed CDW propagation vector.

The softening of phonon modes, as the charge-ordered state is approached from high temperatures, is suppressed by the fluctuations beyond RPA. Since the 
temperature at which the first phonon mode reaches zero energy defines the transition temperature $T_{\textrm{CDW}}$, this means that the charge ordering 
transition temperature in MMA, $T_{\textrm{MMA}}$, is suppressed with respect to the corresponding value in RPA, $T_{\textrm{RPA}}$. 
Choosing the overall strength of the electron-phonon coupling such that $T_{\textrm{MMA}}$ matches 
the experimentally observed value of $T_{\textrm{CDW}}=33.5\,\textrm{K}$, Fig.~\ref{MMA} shows that the corresponding RPA transition occurs already at 
$T_{\textrm{RPA}}\approx300\,\textrm{K}$. Physically, this suppression of the transition temperature can be ascribed to the presence of fluctuations in the 
phonon field.  In the temperature range $T_{\textrm{MMA}}<T<T_{\textrm{RPA}}$, the phonon fluctuations are strong enough to destroy the long range order 
predicted by (mean field) RPA theory. In that regime, the amplitude of the static order parameter vanishes, $\langle\varphi_{\mathbf{Q}}\rangle=0$, but dynamic
fluctuations of the order parameter persist, $\langle|\varphi_{\mathbf{Q}}|^2\rangle>0$~\cite{YoshiyamaEA86,McMillan75}.
The result is a locally-fluctuating short-range ordered state, characterized by the presence of a non-zero order parameter amplitude, without any long-range 
phase coherence~\cite{ChatterjeeEA14}. Due to the non-zero amplitude, the gap in the electronic structure will survive even above the transition temperature 
$T_{\textrm{CDW}}=T_{\textrm{MMA}}$, in the form of a pseudogap phase~\cite{BorisenkoEA09,ChatterjeeEA14}. 
The dynamical fluctuations in this regime may become locally pinned in the presence of defects, creating islands of order up to $T=T_{\textrm{RPA}} \approx 
300\,\textrm{K}$, in agreement  with recent STM and X-ray scattering experiments~\cite{ChatterjeeEA14,ArguelloEA14}. The properties of this regime are 
reminiscent of the features which characterize the pseudogap phase of cuprate high-temperature superconductors~\cite{TimuskEA99}, which have recently
been shown typically to contain charge-ordered phases of their own~\cite{ChangEA12,GhiringhelliEA12}.
Specific similarities are the presence of a pseudogap, Fermi surface arcs, dynamical fluctuations and a corresponding sensitivity to the 
timescales employed by experimental probes, as well as the existence of the crossover to the normal state at $T^*=T_{\textrm{RPA}}$. Note that the existence of 
a crossover temperature $T^*\approx300\,\textrm{K}$, between the pseudogap and disordered regimes predicted here, falls outside of the temperature range that 
has been probed in NbSe$_2$ to date~\cite{BorisenkoEA09,ArguelloEA14}.

\section*{Discussion} 

Although various theories for the formation of charge order in NbSe$_2$ have been able to explain many parts of its diverse experimental signatures, the 
material has so far evaded the simultaneous description of all of its properties by a single consistent theoretical account.
The main features which remain difficult to reconcile with the weak-coupling theories considered so far, include the emergence of CDW order in spite of a lack 
of nesting in the electronic structure, and an anomalously flat electronic susceptibility, the softening of a broad range of phonon modes rather than a sharp 
Kohn anomaly, a discrepancy in the sizes of the CDW gap observed in different electron pockets in spite of them being similarly nested, disagreement about the 
existence, size and particle-hole symmetry of the CDW gap as seen by different experimental probes, and the presence of a pseudogap phase characterized by 
locally fluctuating charge order above the CDW transition temperature. We have shown here that all of these observations can be understood to result from the 
presence of a strong orbital and momentum dependent electron-phonon coupling. 

The combination of both the momentum-dependent coupling and the particular electronic structure of NbSe$_2$ is found to be crucial in the selection of the 
charge ordering vector, while the orbital dependence naturally explains the different behaviours of different Fermi surface pockets. The same dependencies on 
momentum and orbital character are also reflected in the emergence of a particle-hole asymmetric gap, which has a size of about $12\,$meV, centred above $E_F$ 
where it is largely inaccessible to observation by ARPES. These observations put NbSe$_2$ in the regime of strongly-coupled charge-ordered materials, in the 
sense that the specifics of the electron-phonon coupling qualitatively affect the properties of the ordered state. 

The pseudogap phase above $T_{\textrm{CDW}}$ may arise in such a scenario from local fluctuations of the lattice, which destroy the long-range phase coherence 
of the CDW order parameter but not its amplitude. Taking into account higher order phonon fluctuations, we find that indeed the 
mean-field transition temperature is suppressed, leading to a regime above $T_{\textrm{CDW}}$ characterized by the absence of long-range charge order but the 
presence of phonon fluctuations. This agrees with the experimental observation of Fermi arcs and a pseudogap above the transition 
temperature, as well as that of pinned charge order at high temperatures around local defects. All of these signatures of the pseudogap phase are reminiscent 
of similar features seen in the pseudogap phase of cuprate high-temperature superconductors.

In conclusion, while the Peierls mechanism can explain charge ordering in 1D systems entirely in terms of the electronic structure, this is not the natural 
starting point for higher-dimensional materials. Instead, it is necessary to include full information about the dependence of the electron-phonon coupling on 
the momenta and orbital characters of the electronic states involved in the CDW formation. This is evidenced by the present case of NbSe$_2$, a typical 
quasi-2D material without a strongly nested Fermi surface, in which knowledge about all aspects of the electron-phonon coupling prove to be essential in 
explaining the full range of experimental observations. 
 We suggest this consideration to be generically applicable across charge-ordered quasi-2D and quasi-3D 
materials lacking strong nesting features. In particular, the full momentum and orbital dependence of the electron-phonon coupling should be taken into account throughout the class of transition metal dichalcogenides and related material classes, including high-temperature superconductors.

~\\ \indent
{\bf Acknowledgments} \\
The authors thank M. R. Norman for suggesting the approach discussed here, and for numerous insightful discussions.

\bibliographystyle{apsrev4-1}

\end{document}